# Geometric Deep Learning to Identify the Critical 3D Structural Features of the Optic Nerve Head for Glaucoma Diagnosis

**Fabian A. Braeu**[1,2,3], **Alexandre H. Thiéry**[4], **Tin A. Tun**[5,6], **Aiste Kadziauskiene**[7,8], **George Barbastathis**[2,9], **Tin Aung**[3,5,6], and **Michaël J.A. Girard**[1,6,10]

1. Ophthalmic Engineering & Innovation Laboratory, Singapore Eye Research Institute, Singapore National Eye Centre, Singapore
2. Singapore-MIT Alliance for Research and Technology, Singapore
3. Yong Loo Lin School of Medicine, National University of Singapore, Singapore
4. Department of Statistics and Applied Probability, National University of Singapore, Singapore
5. Singapore Eye Research Institute, Singapore National Eye Centre, Singapore
6. Duke-NUS Graduate Medical School, Singapore
7. Clinic of Ears, Nose, Throat and Eye Diseases, Institute of Clinical Medicine, Faculty of Medicine, Vilnius University, Vilnius, Lithuania
8. Center of Eye diseases, Vilnius University Hospital Santaros Klinikos, Vilnius, Lithuania
9. Department of Mechanical Engineering, Massachusetts Institute of Technology, Cambridge, Massachusetts 02139, USA
10. Institute for Molecular and Clinical Ophthalmology, Basel, Switzerland

# Abstract

**Purpose**: The optic nerve head (ONH) undergoes complex and deep 3D morphological changes during the development and progression of glaucoma. Optical coherence tomography (OCT) is the current gold standard to visualize and quantify these changes, however the resulting 3D deep-tissue information has not yet been fully exploited for the diagnosis and prognosis of glaucoma. To this end, we aimed: **(1)** To compare the performance of two relatively recent geometric deep learning techniques in diagnosing glaucoma from a single OCT scan of the ONH; and **(2)** To identify the 3D structural features of the ONH that are critical for the diagnosis of glaucoma.

**Methods:** In this study, we included a total of 2,247 non-glaucoma and 2,259 glaucoma scans from 1,725 subjects. All subjects had their ONHs imaged in 3D with Spectralis OCT. All OCT scans were automatically segmented using deep learning to identify major neural and connective tissues. Each ONH was then represented as a 3D point cloud. We used PointNet and dynamic graph convolutional neural network (DGCNN) to diagnose glaucoma from such 3D ONH point clouds and to identify the critical 3D structural features of the ONH for glaucoma diagnosis.

**Results:** Both the DGCNN (AUC: 0.97±0.01) and PointNet (AUC: 0.95±0.02) were able to accurately detect glaucoma from 3D ONH point clouds. The critical points formed an hourglass pattern with most of them located in the inferior and superior quadrant of the ONH.

**Discussion:** The diagnostic accuracy of both geometric deep learning approaches was excellent. Moreover, we were able to identify the critical 3D structural features of the ONH for glaucoma diagnosis that tremendously improved the transparency and interpretability of our method. Consequently, our approach may have strong potential to be used in clinical applications for the diagnosis and prognosis of a wide range of ophthalmic disorders.

# Introduction

Glaucoma is the leading cause of irreversible blindness affecting about 70 million people worldwide [1-3] and is projected to be 111.8 million in 2040 [4]. It is a multifactorial disease that is characterized by progressive structural damage to the optic nerve head (ONH) – a complex three-dimensional (3D) structure located at the posterior pole of the eye – and loss of retinal ganglion cell axons within the ONH [5, 6]. Optical coherence tomography (OCT), a fast, high-resolution, quantitative, and non-invasive 3D imaging modality, allows the visualization and quantification of such complex 3D morphological changes occurring in glaucomatous ONHs [7-9].

Numerous studies have previously assessed the diagnostic capability of several structural parameters derived from OCT scans, e.g., retinal nerve fiber layer (RNFL) thickness in combination with rim area and cup-to-disc ratio [10, 11], ganglion cell inner plexiform layer thickness (GCL+IPL) [12], macular thickness [13, 14], Bruch's membrane opening - minimum rim width (BMO-MRW) and Bruch's membrane opening (BMO) area [15, 16], and lamina cribrosa (LC) depth and LC curvature index [17, 18]. Nonetheless, Li and Jampel concluded that, from a clinical point of view, OCT was necessary but not sufficient for the detection of glaucoma [19]. A reason for this might be that the above proposed ONH parameters may not fully capture the complex 3D structural signature of the glaucomatous ONH.

More recently, deep learning techniques were used to improve glaucoma diagnosis. Based on the input, glaucoma classification was either performed using: **(1)** traditional ONH parameters (see above) [20-22]; **(2)** 2D OCT B-scans [23-25]; or **(3)** 3D OCT volume scans [26-28]. Most of these studies reported promising results, however the first 2 groups were hampered by the fact that they could not fully exploit the complex 3D information of the ONH while the last overcame this limitation by using the whole raw OCT volume as input to a 3D convolutional neural network (CNN). However, these networks were computationally expensive and required a large amount of GPU memory. The authors were forced to either down-sample the input data that resulted in a loss of small structural features in the OCT volumes [26] or adapt the architecture of the neural network to meet memory constraints [27]. Additionally, the vast amount of redundant data in an OCT volume made it hard for the network to identify and extract important features of the ONH that even led to the development of a complex attention-guided 3D CNN [28].

In this study, we aimed: **(1)** To develop an efficient representation of the complex 3D structure of the ONH as a 3D point cloud, thus significantly reducing the amount of redundant information in an OCT scan; **(2)** To compare the performance of two relatively-recent geometric deep learning techniques (PointNet [29] and dynamic graph CNN or DGCNN [30]) in diagnosing glaucoma from such 3D point cloud representations; **(3)** To identify the 3D structural features of the ONH that are critical for the diagnosis of glaucoma.

## Methods

**Patient Recruitment**

We retrospectively included a total of 1,725 subjects from two cohorts: **(1)** 1,616 subjects (2,247 non-glaucoma and 1,505 glaucoma scans) from Singapore of mixed ethnicity (88% Chinese, 6% Malay, and 6% Indian) and **(2)** 109 subjects (0 non-glaucoma and 754 glaucoma scans) from Lithuania of mixed ethnicity (69% Lithuanians, 13% Poles, 18% Russians). Approximately 45% of the subjects had mild (visual field mean deviation (MD) $\geq$ -6.00 dB), 29% moderate (MD of -6.01 to -12.00 dB), and 26% advanced visual field defects (MD < -12.00 dB) [31]. All subjects gave written informed consent. The study adhered to the tenets of the Declaration of Helsinki and was approved by the institutional review board of the respective institutions. A summary of the patient population is shown in **Table 1**.

Glaucomatous eyes were defined as those with vertical cup-disc ratio (VCDR) > 0.7 and/or neuroretinal rim narrowing with repeatable glaucomatous visual field defects whereas non-glaucomatous eyes were those with an IOP < 21 mmHg, an healthy optic disc with VCDR < 0.5, and normal visual field tests. Subjects with corneal abnormalities that potentially can reduce the quality of the scans and with ONH disorders other than glaucoma were excluded from the studies.

**OCT Imaging**

Subjects had their ONHs imaged in a dark room and if necessary their pupil was dilated with tropicamide 1% solution. All OCT scans (horizontal raster scans) were performed with the same imaging device in both countries (Spectralis, Heidelberg Engineering, Germany) and covered the whole ONH region. Signal averaging was used for all scans. While the number of pixels per A-scan and the number of A-scans per B-scan (slice) were always fixed to 496 and 384 respectively, the number of B-scans varied between 45 and 97. With regards to resolution, the distance between B-scans ranged from 26.8 µm to 79.3 µm, the distance between A-scans ranged from 10.1 µm to 15.1 µm (lateral resolution), and the axial resolution was fixed at 3.9 µm.

**Automated 3D Point Cloud Extraction from OCT Volumes**

To represent the complex 3D structure of an ONH in an efficient way, we developed an algorithm to automatically extract 3D point clouds from raw OCT volume scans. An overview of the single steps is shown in **Figure 1**.

First, to homogenize the dataset, all OCT scans of right eyes were flipped to match a left-eye configuration. We then segmented the raw OCT volumes (**Figure 1a**) using the software REFLECTIVITY (Reflectivity, Abyss Processing Pte Ltd, Singapore) that was developed from advances in AI-based ONH segmentation [32, 33]. More specifically, we classified the following tissue structures of the ONH: **(1)** the RNFL and the prelamina tissue (PLT); **(2)** the GCL+IPL; **(3)** all other retinal layers (ORL); **(4)** the retinal pigment epithelium (RPE) with Bruch's membrane (BM) and the BMO points; **(5)** the choroid; **(6)** the OCT-visible part of the peripapillary sclera including the scleral flange; and **(7)** the OCT-visible part of the LC (**Figure 1b**).

Subsequently, for each B-scan, the anterior and posterior boundaries of each tissue layer were identified. For all points lying on the anterior boundary, a nearest neighbor search to the corresponding posterior boundary was performed to extract a local thickness of the respective tissue layer. Such information was extracted to ultimately improve classification performance. For each ONH, we considered the anterior boundaries of all aforementioned tissues, plus the posterior boundaries of the sclera and LC, in order to represent the final point cloud (**Figure 1c**). Overall, a typical ONH point cloud was represented with approximately 20000 points. All ONH point clouds were aligned. Briefly, based on the extracted BMO points, we calculated the center of BMO and fitted a plane to the BMO points using a least square approach. In a next step, each point cloud was centered at the center of BMO and rotated so that the normal vector of the BMO plane was aligned with the axial direction of the scan. Finally, we performed a cylindrical crop (radius of 1.75 mm and aligned with the axial direction) to further homogenize the data (**Figure 1d, 1e**).

**Geometric Deep Learning on Point Clouds: PointNet**

To distinguish glaucomatous from non-glaucomatous ONHs based on 3D ONH point clouds, we used a slightly modified version of PointNet [29]. Point clouds are an unordered set of points. To make the glaucoma diagnosis independent from the arrangement of points in the input point cloud, PointNet extracted features for each point by multilayer perceptrons (MLPs) that were shared between each point in the point cloud and subsequently aggregated the information by a max pooling layer. Finally, the output of the max pooling layer – often referred to as global feature – was processed by some fully connected layers to get the final classification score (glaucoma vs non-glaucoma). Additionally, the classification should be invariant to rigid transformations of the input point cloud. Therefore, a geometric transformation matrix was predicted by a smaller trainable subnetwork that was directly applied to the coordinates of the input point cloud. A detailed explanation of the architecture can be found in [29].

In this study, we slightly modified the originally proposed architecture: **(1)** we used a max pooling layer of dimension 256; **(2)** each point of our input point cloud had four features, namely, the x-, y-, z-coordinate and the local thickness of the respective tissue layer.

**Geometric Deep Learning on Point Clouds: DGCNN**

One disadvantage of PointNet was that, by design, it was not able to exploit fine local structural features of the input point cloud, such as small changes in curvature. To overcome this deficiency, DGCNN was introduced [30]. The main difference between DGCNN and PointNet was that DGCNN replaced the MLPs by so-called edge convolutional (EdgeConv) layers that were able to capture local structures of the object by dynamically computing a local graph and updating the individual point features based on the current k-nearest neighbors of a point in the feature space (**Figure 1**). Details about the architecture can be found in [30].

Compared to the originally proposed DGCNN, we used a max pooling layer of dimension 256, set the number of k-nearest neighbors to 20 for all EdgeConv layers, and included the local tissue layer thickness as an additional feature to the x-, y-, and z-coordinates of each point.

**Performance Comparison: PointNet vs. DGCNN**

To assess the overall performance of both networks, the input data were split in training (70%), validation (15%), and test (15%) sets, respectively. OCT scans from the same subject were exclusively available in any given set. We performed extensive data augmentation. Specifically, augmentation techniques like random cropping, random rotations of the point cloud, random sampling (i.e. randomly picking a subset of points from the input point cloud), and additive Gaussian noise on the x-, y-, and z-coordinate values were used to enrich the training data. PointNet and DGCNN were trained in a supervised manner on a Nvidia RTX A5000 GPU card until optimum performance was reached in the validation set.

We performed a five-fold cross validation and reported the receiver operating characteristic (ROC) curves and the area under the receiver operating characteristic curves (AUCs) as mean ± standard deviation.

**Identification of Critical 3D Structural Features of the ONH**

The specific architecture of PointNet and DGCNN inherently allowed us to identify important regions of the ONH for the task of glaucoma diagnosis by extracting all points that contributed to the final classification score – the so-called critical points. We extracted the critical points for all ONHs in the test set of the best performing split of the five-fold cross validation study. All points were pooled and visualized as a 3D density map. A high density was obtained when a given point had many neighbors (within a 75 μm radius sphere). Since all ONHs were aligned with respect to specific landmarks (BMO plane and center), such a density map should highlight the critical ONH regions that contribute the most to the diagnosis of glaucoma.

**Sensitivity Study – Contribution of each ONH Tissue Layer**

In this study, we also had an opportunity to question the importance of each individual neural/connective tissue in diagnosing glaucoma. To this end, we applied geometric deep learning (i.e. DGCNN) to each individual tissue layer (represented as a 3D point cloud). We performed a five-fold cross validation and reported AUCs as mean ± standard deviation for each tissue layer.

# Results

**Glaucoma Diagnosis using Geometric Deep Learning: PointNet vs DGCNN**

Both networks were able to accurately detect glaucoma from 3D ONH point clouds. Overall, the DGCNN (AUC: 0.97 ± 0.01) performed slightly better than the PointNet (AUC: 0.95 ± 0.02) and the corresponding ROC curves are shown in **Figure 2**.

**Critical 3D Structural Features of the ONH for Glaucoma Diagnosis**

The critical points used to confirm a glaucoma diagnosis are shown in **Figure 3** for one given glaucomatous ONH from the test set. Interestingly, those points were mostly located in the RNFL at peripapillary region and prelamina layers (~60%), but some were also found in connective tissues such as the sclera and LC. This was true for both methods - PointNet and DGCNN. One difference for the DGCNN was that the critical points were more spread out across the whole ONH.

Additionally, we pooled all critical points from all ONHs (test set) and displayed them as a density map (**Figure 4**). In both cases, the critical points formed an hourglass pattern with points mainly positioned in the inferior and superior quadrant. In addition, in the infero-temporal plane, points were distributed within the neuro-retinal rim, at the center of the LC, and near the LC insertion zone (superior region only). We also found that many of the points were located at the outer edges of the ONH. Finally, the critical points derived from the DGCNN were densely distributed and covered larger regions of the ONH when compared to those derived from the PointNet approach.

**Tissue-specific Glaucoma Diagnosis**

The DGCNN trained on the RNFL+PLT layer alone showed a very good diagnostic performance (AUC: $0.96 \pm 0.02$) that was close to the performance of the DGCNN trained on all layers simultaneously (AUC: $0.97 \pm 0.01$). The diagnostic capability of other ONH tissue layers was lower with the choroid showing the worst performance (AUC: $0.84 \pm 0.03$). Connective tissue layers (LC and/or sclera) obtained a fair score (AUC sclera: $0.90 \pm 0.03$, AUC LC: $0.90 \pm 0.02$). A summary of the five-fold cross validation performance of all analyzed neural and connective tissue layers is shown in **Table 2**.

# Discussion

In this study, we developed a simple yet efficient approach to reduce the vast amount of redundant data in an OCT volume scan of the ONH by using prior human knowledge about the tissue structure of the ONH. We extracted point clouds of the ONH from automatically segmented OCT volume scans and assessed their ability to detect glaucomatous optic neuropathy using two different geometric deep learning methods, namely, PointNet and DGCNN. The classification performance of both networks was excellent with a slight advantage for DGCNN. Furthermore, because of their simplicity, both networks allowed us to identify the 3D structural features of the ONH that are critical for the diagnosis of glaucoma.

We found that the performance of both deep neural networks was very good with an average AUC of $0.95\pm0.02$ and $0.97\pm0.01$ for PointNet and DGCNN, respectively. Although we should keep in mind that AUCs are biased towards a specific dataset, our geometric deep learning approach performed equally or slightly better than previous studies that directly used raw OCT volume scans as an input to 3D deep neural networks with AUCs of: **(1)** $0.94\pm0.04$ with a 3D-CNN [26]; **(2)** $0.97\pm0.01$ with 3D-ResNet [27]; and **(3)** 0.94 with an attention-guided 3D-CNN [28]. Due to the vast amount of data in a raw OCT volume scan, the first 2 studies had to down-sample the OCT volume scan, which resulted in a loss of local structural details. In order to prevent down-sampling, the third study developed a complex attention-guided 3D-CNN that excluded redundant regions of the scan based on 3D-gradient class activation maps (CAMs). These occluded scans (in their original resolution) were then used as inputs to another deep neural network for the final classification task. We opted to use a relatively simple approach. Specifically, we used prior 3D knowledge about ONH morphology to pre-process the OCT volume scans and consequently reduce the amount of redundant data. This allowed us to keep local structures of the ONH and at the same time drastically reduce the amount of input data. Our results strongly suggest that adding human knowledge in the pre-processing step helps deep learning to focus on the important regions of the OCT volume scan.

In our study, DGCNN (AUC: 0.97 ± 0.01) outperformed PointNet (AUC: 0.95 ± 0.02). PointNet was the pioneer for deep learning on point clouds [29]. PointNet, and its extensions [34, 35], treat points largely independent of each other, hence, neglecting their geometrical relationships. In order to capture local geometric structures, DGCNN introduced the so-called EdgeConv layer. It dynamically builds a local graph around each point and is capable of grouping points in both Euclidean and semantic space. This might most likely be the reason for the improved performance of the DGCNN. Therefore, we believe that it is important to accurately capture local structural variations exhibited by the ONH tissues, thus discouraging down-sampling of the input OCT volumes.

Our approach was able to identify and highlight the 3D structural features of the ONH that are critical for the diagnosis of glaucoma. Previous studies also reported important regions of the ONH for glaucoma classification based on CAMs [26, 27] and gradient-weighted CAMs (grad-CAMs) [28]. However, such maps were only reported for individual ONHs and no universal understanding of 3D structural features was provided. In addition, such activation maps have strong limitations. CAMs highly depend on the architecture of the network while grad-CAMs suffer from the gradient saturation problem that can often fail to identify all structural features [36]. In contrast and by design, PointNet and DGCNN highlight critical points for the decision-making process that improves the transparency and interpretability of these models. Overall, this might increase clinical acceptance.

Furthermore, we found that most of the critical points laid in the superior and inferior quadrants of the ONH suggesting that these two quadrants have the highest diagnostic power in glaucoma. This is in accordance with many previous studies that reported significant structural changes of glaucomatous ONHs in these quadrants [10, 15, 37-41], and the presence of disc haemorrhages [42] that could contribute to glaucoma progression [43, 44].

For DGCNN, one might also speculate that the location of the critical points coincides with the location of the central retinal vessel trunk and its branches (CRVT&B) – both evident in **Figure 4** in the en-face and sagittal views. Prior studies reported nasalization of the major retinal blood vessels in glaucoma [45-50]. These studies suggested that the CRVT&B may play an important role in the development and progression of glaucoma by, for example, providing structural support to the weaker ONH neural tissues. Structural changes of the CRVT&B in glaucoma might explain the good correlation between the location of the CRVT&B and the extracted critical points. We also observed a concentration of critical points at the outer boundary of the ONH. A possible explanation might be that the smooth layered tissue structure at this location allows the network to easily extract and compare the thickness of single tissue layers between different ONHs. However, such points might also be artefacts and might highly depend on the design of our networks. Further research would be needed to fully understand their actual contributions.

In this study, we also studied the diagnostic capability of different tissue layers of the ONH with DGCNN and observed that the diagnostic performance of the RNFL+PLT layer (AUC: 0.96 ±0.02) was close to that when all tissues were taken into account (AUC: 0.97 ± 0.01). All other layers showed a significantly reduced performance (see **Table 2**). Currently, the most studied and widely accepted measurement for glaucoma diagnosis and prognosis is the RNFL thickness [10, 11, 13, 14, 51-54]. Other macular and ONH parameters used for glaucoma diagnosis include, but are not limited to: GCL+IPL thickness [12, 55, 56], BMO-MRW and BMO area [16, 57-59], LC depth, and LC curvature index [17, 60]. The 3D shape of the RNFL+PLT layer together with its local thickness can act as a surrogate for most of the aforementioned ONH parameters and therefore could explain its almost excellent performance. This suggests that the 3D structure of the RNFL+PLT layer combined with its local thickness might hold all necessary information of the ONH for the task of glaucoma diagnosis and prognosis. Nevertheless, it should also be emphasized that both the sclera and LC ranked 2[nd] and obtained relatively good scores (AUCs: 0.90±0.03 and 0.90±0.02, respectively). This highlights the fact that glaucoma may not just simply be classified as a neurological disorder, and that ONH connective tissues may be responsible, in part, for the progression and development of glaucoma [61].

In this study, several limitations warrant further discussion. First, our approach was only tested on one OCT device (Spectralis, Heidelberg Engineering, Germany). We were able to process and extract point clouds from OCT scans (all raster, but not diagonal) with varying resolution and field of view, however, in the future, we must make our approach device-agnostic to be universally applicable. Second, we did not include any demographic or clinical information such as age or refractive error as well as any information about glaucoma type, glaucoma severity, and visual field loss. This information will be included in a future study as it might help us, for example, to link local structural changes of the ONH to glaucoma progression. Third, the accuracy of the extracted point cloud to represent local structural features of the ONH depends on the performance of the segmentation algorithm. The herein used segmentation software (Reflectivity, Abyss Processing Pte Ltd, Singapore) was tested and validated on a large cohort of glaucomatous and non-glaucomatous ONHs, however, one should keep in mind that the choice of the segmentation algorithm might influence the diagnostic capability of the extracted point clouds.

In summary, we successfully developed an efficient representation of the complex 3D structure of the ONH by extracting point clouds from automatically segmented OCT volume scans. We demonstrated that deep learning on 3D point clouds of the ONH showed excellent performance for the task of glaucoma classification and revealed new insights into the decision-making process. Therefore, our approach may have strong potential to be used in clinical applications for the detection of various types of ONH disorders.

## Acknowledgment

We acknowledge funding from **(1)** the donors of the National Glaucoma Research, a program of the BrightFocus Foundation, for support of this research (G2021010S [MJAG]); **(2)** SingHealth Duke-NUS Academic Medicine Research Grant (SRDUKAMR21A6 [MJAG]); **(3)** the "Retinal Analytics through Machine learning aiding Physics (RAMP)" project that is supported by the National Research Foundation, Prime Minister's Office, Singapore under its Intra-Create Thematic Grant "Intersection Of Engineering And Health" - NRF2019-THE002-0006 awarded to the Singapore MIT Alliance for Research and Technology (SMART) Centre [MJAG/AT/GB].

# Figures

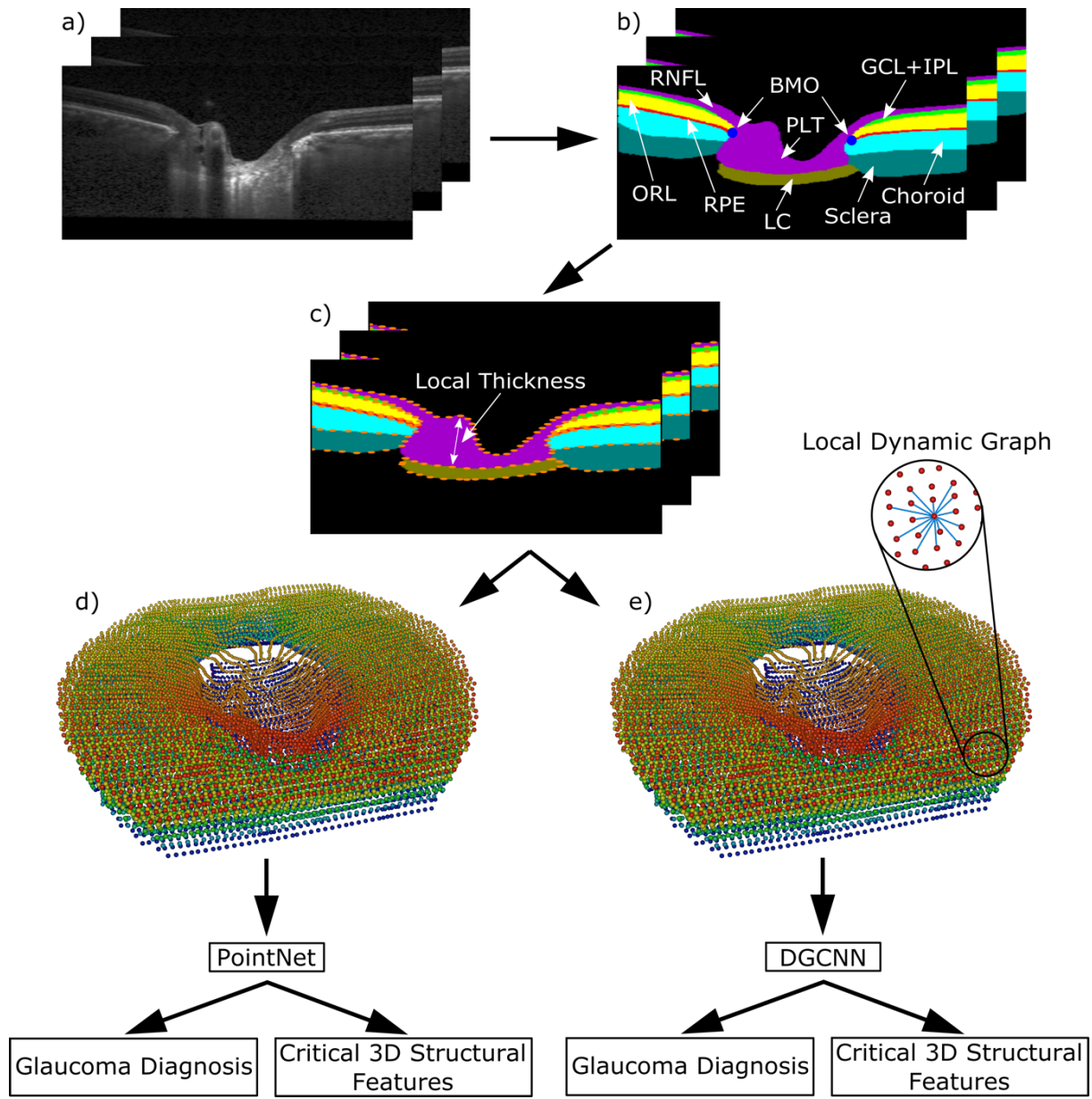

**Figure 1.** Automated extraction of 3D point clouds from raw OCT volumes representing the complex 3D structure of the ONH (a-e). Subsequently, these point clouds were used as input to two deep neural networks (PointNet and DGCNN) for the task of glaucoma classification. Additionally, critical 3D structural features of the ONH for glaucoma diagnosis were inherently identified by the two networks.

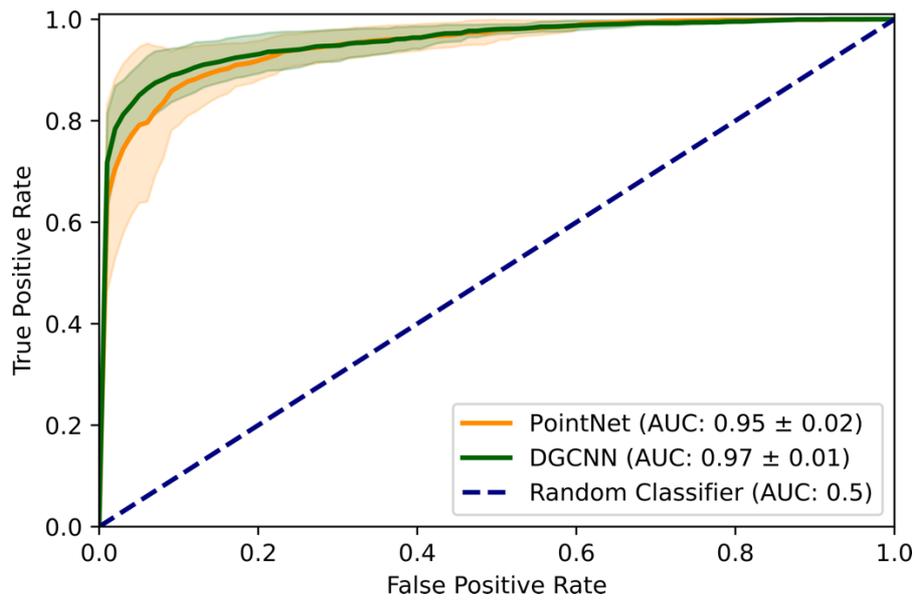

**Figure 2.** Average receiver operating characteristic curves ± standard deviation (shaded area) for the task of glaucoma diagnosis with PointNet (orange) and DGCNN (green).

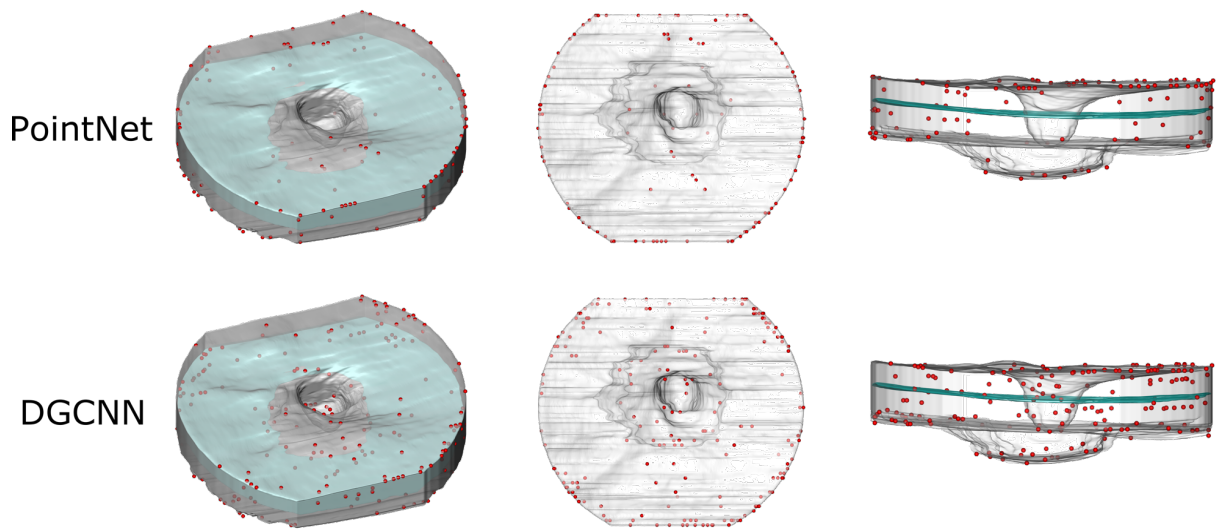

**Figure 3.** Key points of one specific ONH from the test set extracted by PointNet (top row) and DGCNN (bottom row). From left to right: 3d, top, and side view of the ONH. The light blue surface represents the RPE layer. The number of points shown in the figure is smaller than 256 (dimension of the max pooling layer) as some of the points are chosen more than once by the max pooling layer.

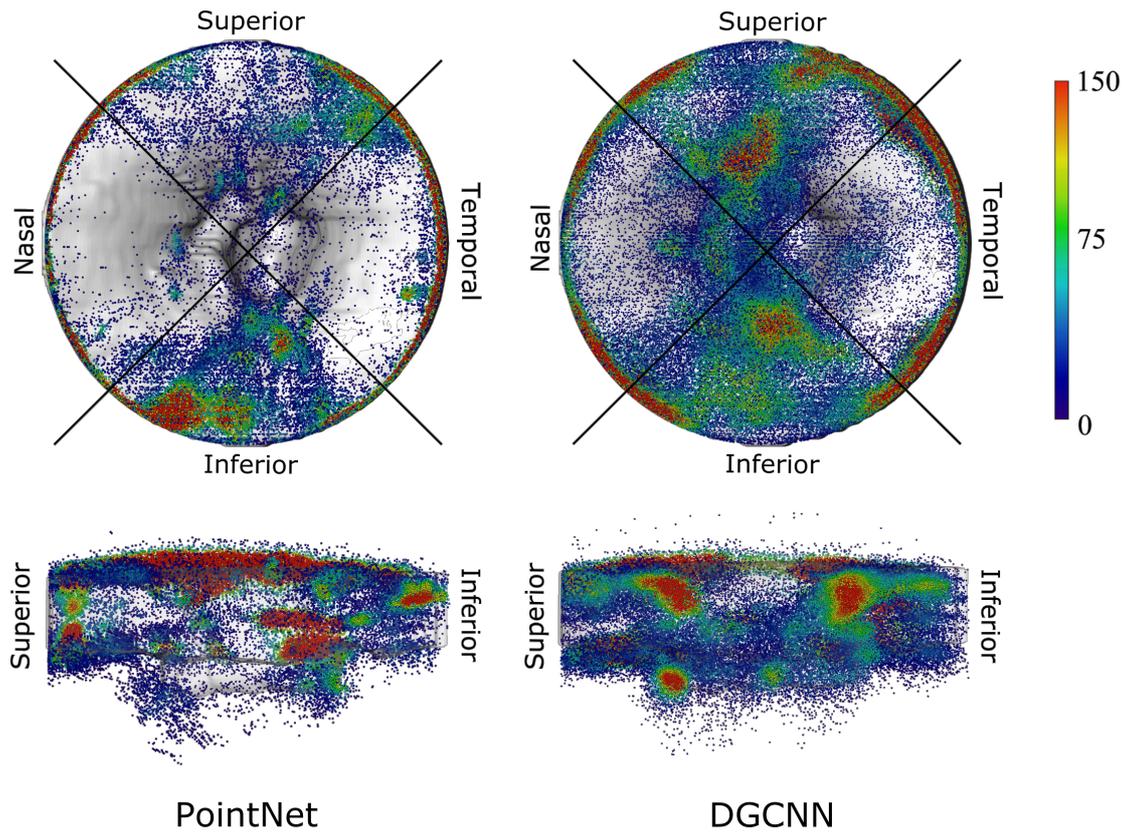

**Figure 4.** En face (top row) and sagittal view (bottom row) of point cloud density maps obtained from PointNet (left) and DGCNN (right) for all ONHs of the best performing test set. The colours indicate the number of neighbouring points within a sphere with a radius of 75 µm.

# Tables

|  | COUNTRY | AGE | SEX (%MALE) | NON-GLAUCOMA SCANS | GLAUCOMA SCANS | TOTAL |
|---|---|---|---|---|---|---|
| **COHORT 1** | Singapore | 62.4±9.0 | 51% | 2,247 | 1,505 | 3,752 |
| **COHORT 2** | Lithuania | 67.3±8.6 | 53% | - | 754 | 754 |
|  | **TOTAL** |  |  | 2,247 | 2,259 | 4,506 |

**Table 1.** Summary of patient populations.

|  | ALL LAYERS | RNFL+PLT | GCL+IPL | ORL | RPE | CHOROID | SCLERA | LC |
|---|---|---|---|---|---|---|---|---|
| **POINTNET** | 0.95 ±0.02 | - | - | - | - | - | - | - |
| **DGCNN** | 0.97 ±0.01 | 0.96 ±0.02 | 0.88 ±0.03 | 0.89 ±0.03 | 0.86 ±0.03 | 0.84 ±0.03 | 0.90 ±0.03 | 0.90 ±0.02 |

**Table 2.** Five-fold cross-validation performance comparison of PointNet and DGCNN trained on all ONH tissue layers and the diagnostic capability of individual ONH layers using DGCNN. Values are average ± standard deviation.